# Diagnosis and Severity Assessment of Ulcerative Colitis using Self Supervised Learning

Venkat Margapuri

*Department of Computer Science*
*Villanova University*
Villanova, PA
vmargapu@villanova.edu

*Abstract*—Ulcerative Colitis (UC) is an incurable inflammatory bowel disease that leads to ulcers along the large intestine and rectum. The increase in the prevalence of UC coupled with gastrointestinal physician shortages stresses the healthcare system and limits the care UC patients receive. A colonoscopy is performed to diagnose UC and assess its severity based on the Mayo Endoscopic Score (MES). The MES ranges between zero and three, wherein zero indicates no inflammation and three indicates that the inflammation is markedly high. Artificial Intelligence (AI)-based neural network models, such as convolutional neural networks (CNNs) are capable of analyzing colonoscopies to diagnose and determine the severity of UC by modeling colonoscopy analysis as a multi-class classification problem. Prior research for AI-based UC diagnosis relies on supervised learning approaches that require large annotated datasets to train the CNNs. However, creating such datasets necessitates that domain experts invest a significant amount of time, rendering the process expensive and challenging. To address the challenge, this research employs self-supervised learning (SSL) frameworks that can efficiently train on unannotated datasets to analyze colonoscopies and, aid in diagnosing UC and its severity. A comparative analysis with supervised learning models shows that SSL frameworks, such as SwAV and SparK outperform supervised learning models on the LIMUC dataset, the largest publicly available annotated dataset of colonoscopy images for UC.

## I. Introduction

Inflammatory bowel disease (IBD) refers to the chronic inflammation of tissues in the digestive tract. Ulcerative Colitis (UC) is a chronic IBD whose etiology remains unknown after decades of research [11] [29]. UC primarily affects the large intestine and is identified by the presence of ulcers in the large intestine and rectum [46]. A study conducted by the Crohn's and Colitis Foundation in 2023 to estimate the incidence, prevalence, and ethnic distribution of physician-diagnosed IBD in the United States found that one in 100 Americans has been diagnosed with IBD and approximately 56,000 new cases are diagnosed each year [27]. The uptick in IBD cases each year warrants a rise in gastrointestinal (GI) physicians. However, the United States Health Resources and Services Administration (HRSA) projects a shortage of 1,630 GI physicians by 2025 and expects the shortage to rise in the next decade. One of the key factors that lead to the shortage is the stress and burnout experienced by the GI physicians on the job [17] [21]. As a consequence, a multitude of patients do not receive timely care for the chronic condition which has the potential to develop into colorectal cancer, if left untreated [24].

UC is diagnosed with a colonoscopy, a test that captures images of the gastrointestinal tract and enables the GI physician to examine the intestinal lining for inflammation, sores, and ulcers. While multiple scoring systems [28] [18] to assess the severity of the disease exist, the Mayo Endoscopic Score (MES) [19] is widely accepted, wherein the severity is classified into one of four categories, Mayo 0 (normal or inactive disease), Mayo 1 (mild disease), Mayo 2 (moderate disease), and Mayo 3 (a severe disease). Artificial Intelligence (AI)-based deep learning (DL) models have demonstrated an excellent ability to perform classification tasks on medical images for clinical diagnosis of UC [39] [40] [45]. In general, DL models help with the reduction in workload for physicians who are faced with an ever-increasing number of medical images to assess, and lower the risk of errors in diagnosis [48] [26]. The majority of DL models are supervised [47] [36], wherein the model is trained on a plethora of annotated image samples. However, publicly available annotated image data sets for colonoscopies are not abundant due to the data confidentiality rules around medical data [1] [12]. Furthermore, the development of annotated colonoscopy image datasets is an endeavor that requires multiple medical experts to dedicate extended amounts of time and energy, making the process tedious and expensive. Self-supervised learning (SSL) is a methodology that develops DL models by learning representations from unlabeled datasets. SSL is proven to be effective for numerous tasks in the medical domain [23] [37], and also in tasks where the data availability is limited. Furthermore, SSL is consistently able to deliver results that are on par with supervised learning, albeit being trained on much smaller annotated datasets. While previous work explores the use of supervised learning on DL models, such as convolutional neural networks (CNN) for

UC diagnosis [3] [6] [2], work using SSL is limited to the state-of-the-art (SOTA) contrastive learning techniques, such as Bootstrap Your Own Latent (BYOL), Momentum Contrast (MoCo), and Swapping Assignments Between Views (SwAV) [35]. Masked autoencoders (MAE) [13] is an SSL technique that has been shown to outperform contrastive learning techniques to train vision transformers, although contrastive learning generally outperforms MAEs for CNNs. However, sparse masked modeling (SparK) [42], a recently proposed MAE-based SSL technique, outperforms the aforementioned SOTA contrastive learning techniques on CNNs. To the best of the authors' knowledge, the MAE-based SSL technique of SparK has not been explored for UC diagnosis using DL models. The current work applies the SOTA contrastive learning techniques, such as BYOL, MoCo, SwAV, and the MAE-based SparK technique for UC diagnosis on the Labeled Images for Ulcerative Colits (LIMUC) dataset [34], the largest publicly available dataset of annotated colonoscopy images for UC. Furthermore, the article benchmarks multiple supervised learning techniques that have been used in prior medical imaging research, such as Google's InceptionV4 [41], Microsoft's ResNet-50 [15], and Oxford's VGG-19 [38] for UC diagnosis. The experiment conducted on the LIMUC dataset demonstrates the feasibility of SSL techniques over supervised learning techniques by achieving superior performance with fewer annotated images.

The key contributions of the paper are:

1) The application of SparK [42], an MAE-based SSL technique, and contrastive learning-based SSL techniques, such as BYOL [10], MoCo [14], and SwAV [7], for the diagnosis of UC on the LIMUC dataset.
2) Benchmarking of supervised learning models, such as Google's InceptionV4 [41], Microsoft's ResNet-50 [15], and Oxford's VGG-19 [38] on the LIMUC dataset.
3) Elaborate experiment and comparative analysis to show that SSL outperforms supervised learning techniques in the absence of large annotated image datasets for the diagnosis of UC and determining its severity.

## II. RELATED WORK

Pretext tasks play a key role in training models using SSL. They refer to tasks wherein the neural network model is trained to learn features of the image dataset by predicting psuedo-labels obtained from applying different transformations to the images. Pretext tasks may be common across multiple downstream tasks. Commonly used pretext tasks include surrogate class prediction [9], relative position prediction [8], image transformation prediction, and solving jigsaw puzzles [33]. In the realm of SSL, He et al. [14] proposed MoCo that builds a moving-average encoder by leveraging a dynamic dictionary with a queue to store feature vectors, ensuring that feature vectors are stored and not computed multiple times. Grill et al. [10] introduced the BYOL architecture that comprises online and target neural network models, each with a different augmented view of an image. The online neural network is trained to predict the target neural network representation of the image. The target neural network is updated with a slow-moving average of the online network. Caron et al. [7] proposed the SwAV architecture, wherein the principle was to cluster different augmentations of the same image instead of performing pair-wise feature comparisons on the augmentations of the image, typical of contrastive learning techniques. He et al. [13] proposed MAE to apply SSL techniques on vision transformers. MAE masks a portion of the image and applies an encoder on the visible patches of the image to generate encoded tokens. The encoded and masked tokens are processed using a decoder to reconstruct the original image. The critical works that inspired the choice of the dataset and frameworks for current work are highlighted next.

Tian et al. [42], proposed sparse masked modeling (SparK), a hierarchical decoder to reconstruct images from multiscale encoded features. It can be used on any CNN without needing backbone modifications. Validations on ResNet and ConvNext models demonstrated that it surpassed contrastive learning techniques for CNNs.

Turan and Durmus [43] explored the use of synthetic images to train a classification model based on MES to assess UC from colonoscopy images. They developed UC-NfNet, an automated classification method, that outperformed other SOTA techniques such as convolutional vision transformers, swin transformer, InceptionV4, and ResNets. The use of synthetic image datasets to train the CNN is a key contribution, considering the lack of publicly available datasets for UC classification.

Pyatha, Xu, and Ali [35] performed vision transformer-based SSL for UC grading using colonoscopy images on the LIMUC dataset. They used the combination of self-supervised MoCo with the Swin Transformer to develop a UC classifier based on MES that outperformed ResNet-50, a supervised learning technique.

Wolf et al. [48] demonstrated the feasibility of SSL for the classification of computed tomography (CT) scan images using the Lung Image Database Consortium and Image Database Resource Initiative (LIDC-IDRI) dataset that consists of CT scans from 1,010 patients. The idea is to pre-train classification models using SSL techniques, such as MoCo and MAE, and leverage the model for the downstream classification tasks of COVID-19, OrgMNIST, and brain hemorrhage.

Bhambhvani and Zamora [3] developed classification models for UC using different variants of ResNet, such as 50 and 101. The dataset used for the experiment is the Hyper-Kvasir [4] dataset, and the images are classified based on the MES.

Byrne et al. [6], compiled a dataset of colonoscopy images by collecting colonoscopy videos of 134 UC patients comprising 1,550,030 frames at the Asian Institute of Gastroenterology (AIG). The images were used to predict the MES and Ulcerative Colitis Endoscopic Index of Severity (UCEIS) score using the augmented

EfficientNetB3 architecture that they labeled Section-based Disease Assessment (SDA).

Becker et al. [2] developed a model for UC classification based on the Mayo Clinic Endoscopic Subscore (MCES) using sigmoidoscopy from three different clinical study trials, Eucalyptus, Hickory, and Laurel. The overall classifier comprised the Quality Control (QC) model and Ulcerative Colitis Scoring (UCS) model. The QC model processed each frame of the sigmoidoscopy videos and discriminated between readable and non-readable frames. The readable frames were used to predict UC severity using the UCS model.

## III. SELF-SUPERVISED LEARNING

SSL refers to the technique of training neural network models to learn image features where the training data is not labeled (annotated). The neural networks are provided with pseudo labels that indicate the different image transformations applied to the images. In the absence of well-defined labels, the neural network groups different images in the dataset based on the similarity of pseudo labels associated with the images. Four different frameworks namely, BYOL, MoCo, SwAV, and SparK are used as part of the experiment to diagnose UC using SSL. Each of the architectures is described further.

### A. BYOL

The basic principle of BYOL is to group augmented image samples that are similar to each other. One of the key benefits of BYOL is that it avoids the collapsed representation problem. Collapsed representation is the state wherein a network trained only on similar pairs of image samples learns a constant function that sets the loss to a constant value, such as zero. As a result, no discriminative features are learned and the network remains unfit for fine-tuning on a different downstream task. The working of BYOL is described as follows:

1) **Target and Online Networks:** The architecture comprises two image encoders, such as ResNet-50, with the same architecture. One of them is the 'target' network that is initialized with random parameters, and the other is the trainable 'online' network.
2) **Image Generation:** An input image 'I' is run through an image augmentation pipeline to generate 'I1' and 'I2', two independent stochastically augmented versions of 'I'.
3) **Vector Encoding:** The images 'I1' and 'I2' are processed using the 'target' and 'online' networks respectively to extract the vector encodings for each of the augmented views of 'I'. The vector encodings are reduced to a low dimensional latent space, such as 256, if they are initially generated in a high dimensional vector space.
4) **Prediction:** The 'online' vector representation is used to predict the 'target' vector representation by minimizing the distance between the two vector encodings using the normalized mean squared error loss function.
5) **Updates:** The 'target' network is updated at the end of each training step as the exponential moving average of the parameters of the 'online' network.

### B. MoCo

MoCo is a technique that is based on the principle of matching queries to keys wherein key and query are vector encodings generated by passing augmented versions of the image through two independent encoders, such as ResNet-50, that are architecturally similar. MoCo constructs a dictionary, known as the feature queue, which operates as a queue that keeps a history of the encoded keys. The feature queue is used to create positive and negative pairs of images. A positive pair refers to the instance where the query matches the key. All the keys that don't correspond to the query are assumed to form a negative pair with the query.

The key benefit of maintaining the dictionary as a queue is its flexible size, wherein the queue can hold multiple mini-batches of keys at once. As a new mini-batch of keys is enqueued, the oldest mini-batch is dequeued. However, the key pitfall of the feature queue-based encoder is that it isn't conducive to backpropagation owing to its large size. As a workaround, the MoCo framework leaves the parameters of the key encoder frozen during backpropagation, and updates only backpropagates the gradient to the query encoder. A momentum-based average of the query encoder is computed and used to update the parameters of the key encoder. It is computed as $\vartheta_k \leftarrow m\vartheta_k + (1 - m)\vartheta_q$ where $\vartheta_k$ and $\vartheta_q$ are the parameters of the key and query encoders respectively, and m is the momentum whose value is maintained close to one. The Contrastive Loss function which computes the similarity between the query and key is used to compute the loss during training. It is formulated as $L^{\text{self}} = -log \frac{exp(q.k^+/\tau)}{\sum_k exp(q.k/\tau)}$ where k, $k^+$, q, and $\tau$ refer to key, positive key, query, and temperature respectively. The lower the temperature, the higher the confidence of the model, and vice versa.

### C. SwAV

SwAV is an online clustering technique that works well with small and large batch sizes without the need for a momentum encoder or memory store. The standout feature of SwAV that makes it unique from most SSL techniques is Multi-Crop Augmentation (MCA). Unlike other SSL techniques, such as MoCo and BYOL, that only create one augmented view of the original image, SwAV creates multiple augmented views of the original image by applying different image transformations. Since it is computationally expensive to create multiple crops that are high resolution, SwAV creates a mix of high-resolution and low-resolution images to keep the computational cost reasonable. The augmented image views are passed in pairs through an encoder, such as ResNet-50, to generate high-dimensional encoded feature vectors. The feature vectors are normalized by projecting them onto the unit sphere to ensure consistency. The normalized feature vectors are matched with prototype vectors that are characteristic of clustering algorithms. Prototype vectors define the space of possible distinction between the learnable feature space, and may also be understood as a low-dimensional projection of the dataset. The mapping of feature vectors to the prototype vectors is done by computing the similarity between them resulting in a clustering assignment. The intuition behind creating the clusters is that "different augmented views (crops) taken from the same image should produce similar feature vectors." Therefore, the output vectors for the image crops are swapped, i.e. crop 1 is assigned output vector 2 and vice versa. The model is trained to predict the output vector of a different augmented view as the target. The loss function used by SwAV is formulated as $L(Z_t, Z_s) = -(\sum_k q_s^{(k)} \log P_t^k + \sum_k q_t^{(k)} \log P_s^k)$ where $Z_t$ and $Z_s$ refer to the feature vectors from different augmentations of the same image, $q_s$ and $q_t$ refer to the output vectors obtained after clustering, and $P_t$ and $P_s$ refer to the probability obtained by taking the softmax of the dot product between the image feature vectors and prototype vectors.

### D. SparK

SparK is a technique that is primarily based on MAE in the natural language processing (NLP) space where models are pre-trained by dividing the image into patches, masking some of the patches, and predicting the masked patch to reconstruct the original image. While MAEs work well for vision transformers, their performance on CNNs has been moderate. SparK is an improvement over the current state of MAE, where it has been shown to outperform SOTA SSL techniques for CNNs [42]. The architecture is explained as follows:

1) **Image Curation:** Given an image dataset I = { $I_1$, $I_2$, $I_3$,... } each image is divided into multiple non-overlapping square patches. Each of the patches is masked independently with a given probability known as "mask ratio."
2) **Image Encoder:** The model consists of an encoder, such as ResNet-50, to perform sparse convolutions wherein computations are performed only when the center element of the sliding window kernel is not empty. It ensures that computations are not performed on masked regions and that feature vectors (encodings) are constructed only where the image is not masked.
3) **Image Decoder:** The feature vectors from the encoder are passed in four different resolutions to a decoder whose architecture is similar to the U-Net [16] design, containing three blocks of upsampling layers. Any empty parts in the feature vectors are filled with learnable mask embeddings M as they are input to the decoder. Furthermore, a projection layer is added between the encoder and decoder for all computed resolutions in case they have different network widths.
4) **Image Reconstruction:** The head module applied after the decoder with two upsampling layers reconstructs the image in the original resolution. Per-patch normalized pixels are chosen as targets, with L2 being the loss function that only computes errors on the masked areas of the images.

## IV. DATASET AND METHOD

### A. Dataset

The image dataset leveraged for the experiment is the LIMUC dataset which comprises 11,276 images from 564 patients, and 1,043 colonoscopy procedures. While the size of the LIMUC dataset may not be considered large for experiments in other domains, it is worth noting that medical data is classified and the LIMUC dataset is considered large in the medical domain. The images are compiled from patients who underwent colonoscopy for UC between December 2011 and July 2019 in the Department of Gastroenterology at Marmara University School of Medicine. The compilation of the dataset includes two experienced gastroenterologists who blind-reviewed and classified all the images according to the MES. The images that lacked classification consensus are inspected independently by a third gastroenterologist, and the final label is assigned by majority voting. Table I shows the breakdown of the number of images classified into each of the four different MES categories.

TABLE I
LIMUC DATASET IMAGE CLASSIFICATION

| MES Category | Image Count |
|---|---|
| Mayo 0 | 6105 (54.14%) |
| Mayo 1 | 3052 (27.70%) |
| Mayo 2 | 1254 (11.12%) |
| Mayo 3 | 865 (7.67%) |

The MES classification is indicative of the severity of UC, and sample images from the LIMUC dataset for each of the MES categories are shown in Fig. 1. It can be visually observed that the score of Mayo 0 indicates no inflammation, and the inflammation progressively worsens as the MES increases.

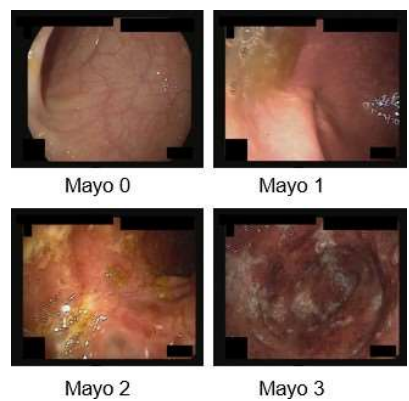

Fig. 1. Colonoscopy images with different MES categorizations

### B. Research Questions

The experiment described in the paper for UC diagnosis and severity assessment aims to answer the following research questions (RQ):

- RQ-1: How do SSL techniques compare with supervised learning techniques?
- RQ-2: How does each of the SSL techniques compare with the others?
- RQ-3: How does the size of the dataset affect the performance of the supervised learning and SSL techniques?
- RQ-4: How does class imbalance within the dataset affect the performance of the supervised learning and SSL techniques?

### C. Experiment

The experiment to answer the RQ in IV-B involves conducting experiments using the LIMUC dataset on supervised learning and SSL models. The experiments are described as follows:

- Supervised Learning: Google's InceptionV4, Microsoft's ResNet-50, and Oxford's VGG-19 models are chosen for the experiment due to their varied architectures, and success in the field of medical imaging. The LIMUC dataset, shown in Table I, is split into Train, Validation, and Test datasets, each containing 60%, 20%, and 20% of the data respectively. As a means to understand the impact of the size of the training dataset on the performance of the supervised models, they are independently trained on 100%, 50%, and 25% of the data from the Train dataset. It is to be noted that using a part of the Train dataset still preserves images from each MES category at the same rate that they originally appear in the Train dataset. The supervised models are pre-trained on the ImageNet dataset, and augmented to finetune on the Train dataset. Three fully connected layers are added to the base architectures, followed by a softmax layer for classification. All the pre-trained layers in the base architectures are frozen, training only the three fully connected layers. Table II shows the results of evaluating each of the supervised models on the Test dataset upon being trained on different splits of the Train dataset.

TABLE II
PERFORMANCE OF SUPERVISED LEARNING MODELS ON THE TEST DATASET UPON BEING TRAINED ON DIFFERENT SIZES OF THE TRAIN DATASET

| training with 100% data from Train dataset | | | | |
|---|---|---|---|---|
| Method | Accuracy | Precision | Recall | F1-Score |
| Inception-V4 | 58.1 | 63.3 | 53.7 | 58.2 |
| ResNet-50 | 73.2 | 70.1 | 71.7 | 70.9 |
| VGG-19 | 66.7 | 68.2 | 66.7 | 67.4 |
| training with 50% data from Train dataset | | | | |
| Method | Accuracy | Precision | Recall | F1-Score |
| Inception-V4 | 53.7 | 62.2 | 49.9 | 55.4 |
| ResNet-50 | 68.9 | 70.2 | 66.9 | 68.5 |
| VGG-19 | 65.4 | 66.1 | 64.8 | 65.4 |
| training with 25% data from Train dataset | | | | |
| Method | Accuracy | Precision | Recall | F1-Score |
| Inception-V4 | 50.1 | 60.9 | 44.8 | 51.6 |
| ResNet-50 | 64.6 | 65.4 | 66.2 | 65.6 |
| VGG-19 | 63.7 | 65.2 | 62.8 | 63.9 |

The metrics of accuracy, precision, recall, and F1-score are used to evaluate the performance of the model. They are briefly described as they relate to multi-class classification, and expressed as percentages for better readability wherever observed in the article.

- **Accuracy:** Accuracy is defined as the ratio of the number of correct predictions to the total number of predictions made on the dataset. It is computed as $\frac{true\ positives}{(true positives + false positives + true negatives + false negatives)}$
- **Precision:** Precision for a class $i$ is the ratio of the correctly predicted positives for class $i$ to the total number of predicted positives for class $i$. It is computed as $\frac{true\ positives}{true\ positives\ +\ false\ positives}$.
- **Recall:** Recall for a class $i$ is the ratio of the correctly predicted positives for class $i$ to all the data samples that belong to class $i$. It is computed as $\frac{true positives}{true positives + false positives}$.
- **F1-score:** F1-score is given by the harmonic mean of precision and recall and is computed as $2 * \frac{precision\ *\ recall}{precision\ +\ recall}$

In a nutshell, precision and recall provide insight into the cost of false positives and false negatives affecting the UC diagnoses. For instance, false positives may lead to unnecessary treatments and false negatives may lead to missed diagnoses. Either scenario is a traumatic experience for the patients and needs to be avoided.

**Note:** The standard/macro precision, recall, and F1-score are reported as part of the experiment. While the "micro" versions are deemed more robust to class imbalance, the experiment accounts for class imbalance already, as described in the "Addressing Class Imbalance" section below. In an attempt to ensure that the metrics are not skewed by computation, the "standard/macro" version is reported.

- Self-supervised Learning: The experiment using SSL comprises two steps:

  1) **Pre-training:** The SSL frameworks described in section III are initially trained on a majority of the images from the LIMUC dataset without labels.
  2) **Fine-tuning:** The SSL model trained without labels is fine-tuned (trained) on a small portion of the images from the LIMUC dataset with labels.

In order to train the SSL frameworks, the data from the LIMUC dataset is split into three datasets, Pre-train, Fine-tune, and Test. 50%, 30%, and 20% of the data are allocated to Pre-train, Fine-tune, and Test datasets respectively. Table III shows the number of images per MES category assigned to each of the three datasets.

TABLE III
IMAGES PER MES CATEGORY IN PRE-TRAIN, FINE-TUNE, AND TEST SETS

| | Pre-train | Fine-tune | Test |
|---|---|---|---|
| Mayo 0 | 3053 | 1832 | 1220 |
| Mayo 1 | 1526 | 916 | 610 |
| Mayo 2 | 627 | 376 | 251 |
| Mayo 3 | 433 | 260 | 172 |

Each of the SSL frameworks described in section III is pre-trained without labels on the Pre-train dataset using the backbone (feature extractor) of ResNet-50 for its superior performance amongst the experimented supervised learning models. In order to evaluate the impact of the size of the dataset on the performance of the SSL frameworks, each of the four SSL models is pre-trained using two different data configurations of the Pre-train dataset, 100% and 50%, developing two pre-trained models per SSL framework. Each of the pre-trained models is then trained with labels on three different data configurations of the Fine-tune dataset, 100%, 50%, and 25%. Table IV and Table V show the performance of the SSL models when trained on 100% and 50% of the Pre-train dataset with different data configurations of the Fine-tune dataset, respectively. Similar to the supervised learning models, the SSL models are evaluated on the metrics of accuracy, precision, recall, and F1-score.

TABLE IV
FINETUNED MODEL PERFORMANCE UPON PRE-TRAINING WITH 100% OF THE DATA FROM THE PRE-TRAIN DATASET

| Method | Accuracy | Precision | Recall | F1-Score |
|---|---|---|---|---|
| \multicolumn{5}{c}{finetuning with 100% data from Fine-tune dataset} |
| BYOL | 65.8 | 66.3 | 66.8 | 66.5 |
| MoCo | 68.8 | 67.1 | 66.7 | 66.9 |
| SwAV | 70.1 | 69.1 | 71.7 | 70.4 |
| SparK | 73.9 | 70.8 | 75.6 | 73.1 |
| \multicolumn{5}{c}{finetuning with 50% data from Fine-tune dataset} |
| BYOL | 63.6 | 64.9 | 66.1 | 65.4 |
| MoCo | 62.2 | 66.8 | 65.4 | 66.1 |
| SwAV | 66.3 | 66.7 | 70.3 | 68.4 |
| SparK | 70.2 | 68.1 | 73.9 | 70.9 |
| \multicolumn{5}{c}{finetuning with 25% data from Fine-tune dataset} |
| BYOL | 59.3 | 61.5 | 62.4 | 61.8 |
| MoCo | 59.8 | 60.1 | 64.2 | 62.1 |
| SwAV | 65.8 | 66.4 | 71.5 | 68.8 |
| SparK | 69.6 | 69.3 | 71.1 | 70.2 |

TABLE V
FINETUNED MODEL PERFORMANCE UPON PRE-TRAINING WITH 50% OF THE DATA FROM THE PRE-TRAIN DATASET

| Method | Accuracy | Precision | Recall | F1-Score |
|---|---|---|---|---|
| \multicolumn{5}{c}{finetuning with 100% data from Fine-tune dataset} |
| BYOL | 65.1 | 64.3 | 65.6 | 65.1 |
| MoCo | 67.4 | 64.6 | 68.3 | 65.2 |
| SwAV | 68.5 | 66.7 | 70.9 | 68.7 |
| SparK | 69.9 | 67.3 | 72.5 | 69.8 |
| \multicolumn{5}{c}{finetuning with 50% data from Fine-tune dataset} |
| BYOL | 62.7 | 63.9 | 65.2 | 64.7 |
| MoCo | 61.4 | 63.4 | 64.9 | 64.2 |
| SwAV | 66.1 | 65.2 | 69.1 | 67.1 |
| SparK | 68.3 | 66.9 | 71.4 | 69.1 |
| \multicolumn{5}{c}{finetuning with 25% data from Fine-tune dataset} |
| BYOL | 60.8 | 62.1 | 64.6 | 63.5 |
| MoCo | 60.5 | 61.9 | 63.6 | 62.7 |
| SwAV | 64.6 | 63.7 | 69 | 66.3 |
| SparK | 66 | 66.6 | 70.9 | 68.6 |

- Addressing Class Imbalance: A key criterion that the experiment does not take into account yet is class imbalance. From Table I, one may observe that the majority class, Mayo 0, accounts for 54.14% of the images, whereas the largest minority class, Mayo 3, accounts for a mere 7.67% of the images. The two other minority classes, Mayo 1 and Mayo 2, comprise 27.70% and 11.12% of the images respectively. The ill effects of class imbalance in deep learning are well studied and affect both supervised and SSL frameworks [5] [30] [32]. Techniques such as resampling, class weighting, and data augmentation are used in deep learning to address class imbalance. The application of resampling or data augmentation results in the size of the dataset being altered due to an increase in the number of images for the minority class, whereas class weighting augments the loss function to penalize the misclassifications of the minority class more heavily than the majority class. The data in the LIMUC dataset is classified by experts, and augmented images cannot be trusted with their labels unless classified by a domain expert. However, it is a laborious and expensive process. Therefore, the application of the class weighting technique to address class imbalance is more feasible in the current scenario. Class weighting is applied to the loss functions of the supervised learning and SSL models using the Scikit-learn library [22]. The weight, $w_i$, for each class $i$ in the dataset is computed as $\frac{n_{images}}{n_{classes} * n_i}$ where $n_{images}$ is the number of images in the dataset, $n_{classes}$ is the number of unique classes in the dataset, and $n_i$ is the number of samples in class $i$. The experiment to determine the impact of class imbalance is conducted on the supervised and SSL models. Class weighting is applied to the supervised models trained on 100% of the Train dataset, and SSL models pre-trained with 100% of the data from the Pre-train dataset and finetuned with 100% of the data from the Fine-tune dataset. The results for both supervised and SSL models are shown in Table VI.

TABLE VI
PERFORMANCE OF SUPERVISED AND SELF SUPERVISED LEARNING MODELS UPON ACCOUNTING FOR CLASS IMBALANCE

| Method | Accuracy | Precision | Recall | F1-Score |
|---|---|---|---|---|
| \multicolumn{5}{c}{Supervised Learning Models} |
| Inception-V4 | 59.1 | 63.9 | 55.3 | 59.2 |
| ResNet-50 | 75.9 | 71.9 | 73.1 | 72.5 |
| VGG-19 | 68.5 | 66.7 | 70.9 | 68.7 |
| \multicolumn{5}{c}{Pre-trained Self Supervised Learning Models} |
| BYOL | 66.8 | 67.1 | 68.2 | 67.7 |
| MoCo | 69.4 | 68.6 | 68.7 | 68.9 |
| SwAV | 77.9 | 74.6 | 75.7 | 75.5 |
| SparK | 79.8 | 76.4 | 79.3 | 77.5 |

D. Evaluation

The results from the experiment are used to answer the four RQ mentioned in IV-B.

1) How do SSL techniques compare with supervised learning techniques?
   Considering the class imbalance in the datasets that are employed on the supervised and SSL techniques, the key metric that aids in the comparison of SSL techniques with the supervised learning techniques is the F1-score [20]. Furthermore, it is worth noting that F1-score considers the metrics of precision and recall in its derivation. Among the supervised learning models (Table II), ResNet-50 performs the best with an F1-score of 70.9%, followed closely by VGG-19 with an F1-score of 67.4%. Among the SSL models (Table IV and Table V), the MAE-based technique of SparK performs the best with an F1-score of 73.1% followed by the contrastive learning technique of SwAV with an F1-score of 70.4%. It is important to note that the supervised and SSL models perform their best when trained on 100% of their respective datasets. The performance metrics illustrate that SSL frameworks offer similar (or better) performance compared with the supervised learning models. Furthermore, they don't require the same volume of labeled data, making them more feasible for UC diagnosis and severity assessment.

2) How does each of the SSL techniques compare with the others?
   Comparing the results of the SSL models with each other (Table IV and Table V), it is observed that the SSL models achieve their best accuracy and F1-score metrics when trained on 100% of the Pretrain and Fine-tune datasets respectively. The MAE-based SparK technique, with an accuracy of 73.9% and F1-score of 73.1%, outperforms the contrastive learning approaches of BYOL, MoCo, and SwAV. Among the contrastive learning models, SwAV, with an F1-score of 70.4% outperforms both MoCo and BYOL with F1-scores of 66.9% and 66.5% respectively. The superior performance of SwAV can be attributed to the MCA of SwAV wherein multiple augmentations at different resolutions are created, thereby enabling the framework to learn features at different scales. In contrast, the contrastive learning techniques of BYOL and MoCo rely on only one augmented view of the original image. Like SwAV, the SparK technique comprises a decoder that reconstructs images from multiscale encoded features. Therefore, it is inferred that SSL frameworks that consider multiple augmentations of the original image at different resolutions perform better than those that don't.

3) How does the size of the dataset affect the performance of the supervised learning and SSL techniques?
   The impact of dataset size is evident in the performance of both the supervised learning and SSL frameworks. From the results in Table II pertinent to the supervised learning models, it is observed that each of the models delivers the best metrics when trained on 100% of the Train dataset, and the performance gets progressively worse as the size of the training dataset decreases. Likewise, from the results in Table IV and Table V, it is observed that the SSL frameworks perform best when pre-trained and fine-tuned with 100% of the Pre-train and Fine-tune datasets. Observing further, the performance of the supervised models drops sharply compared to the SSL models as the size of the dataset decreases. For instance, the best-performing supervised learning model, ResNet-50, exhibits F1-scores of 70.9%, 68.5%, and 65.6% at 100%, 50%, and 25% of training data respectively (Table II). A decrease of 2.4% and 3.2% are observed for every 50% decrease in training data. On the other hand, the best-performing SSL framework, SparK, exhibits F1-scores of 73.1%, 70.9%, and 70.2% when pre-trained on 100% of the Pre-train dataset and fine-tuned on 100%, 50%, and 25% of the Fine-Tune dataset respectively (Table IV). A decrease 2.2% and 0.7% are observed for every 50% decrease in the finetuning data. Similar observations are made in reference to the other supervised and SSL models, effectively demonstrating the robustness of SSL frameworks compared with supervised models when trained on varying data sizes.

4) How does class imbalance within the dataset affect the performance of the supervised learning and SSL techniques?
   From the results in Table VI, it is observed that class imbalance within the dataset has a notable impact on the performance of both the supervised learning and SSL frameworks. The performance metrics of all the models, supervised and self supervised, have improved; ResNet-50 and SparK continue to outperform the other supervised learning and SSL models, respectively. Upon accounting for class imbalance, two of the SSL frameworks, SparK and SwAV with F1-scores of 77.5% and 75.5%, outperform the best supervised learning model, ResNet-50, with an F1-score of 72.5%. While it may appear that the increase in the F1-scores for supervised learning and SSL frameworks is only about 3% to 5% when accounted for class imbalance, it is important to note that such minor improvements in performance are significant in the diagnosis of UC for the following reasons backed by prior research.

- **Trust and Adoption:** Healthcare providers and patients are more likely to trust applications that improve in performance over time [31].
- **Impact on Unusual Conditions:** For rare/unusual conditions, even a 1% increase in performance can significantly enhance the model's ability to detect cases that might otherwise be missed [44].
- **Improved Generalization:** An increase in model

performance indicates the model's ability to generalize better to unseen data [25].

## V. CONCLUSION AND FUTURE WORK

UC is an incurable autoimmune condition that gravely affects the quality of life of those afflicted. While AI systems are capable of diagnosing the severity of UC from colonoscopy images, the majority of earlier research with tangible results focuses on leveraging supervised learning frameworks that require large labeled datasets. The current work demonstrates the feasibility of SSL frameworks which work efficiently even with small labeled datasets. Furthermore, it establishes benchmarks for different SSL frameworks to diagnose the severity of UC.

In the future, the feasibility of SSL frameworks on different colonoscopy datasets will be investigated. Furthermore, the current work does not focus on explainable AI techniques to interpret the model behavior. In the future, explainable AI techniques, such as Gradient Class Activation Mapping, will be applied to further improve the performance of the models.